\documentclass[11pt]{article}
\usepackage{axodraw}
\usepackage{epsfig}
\usepackage{amsfonts}
\usepackage{amsmath}
\usepackage{mathabx}
\usepackage{bbm}
\input pix.sty

\renewcommand{\eq}{eq.~}
\renewcommand{\eqs}{eqs.~}
\renewcommand{\se}{sec.~}

\renewcommand{\fig}{fig.~}
\renewcommand{\figs}{figs.~}

\newcommand{\Nf}{N_{\rm f}}
\newcommand{\Nc}{N_{\rm c}}

\newcommand{\rmO}{{\mathcal{O}}}

\def\lsi{\raise0.3ex\hbox{$<$\kern-0.75em\raise-1.1ex\hbox{$\sim$}}}
\def\gsi{\raise0.3ex\hbox{$>$\kern-0.75em\raise-1.1ex\hbox{$\sim$}}}
\newcommand{\lsim}{\mathop{\lsi}}
\newcommand{\gsim}{\mathop{\gsi}}
\renewcommand{\lsim}{\lesssim}
\renewcommand{\gsim}{\gtrsim}

\newcommand{\nB}{n_\rmii{B}}

 \renewcommand{\nB}[1]{n_\rmii{B{#1}}}
\newcommand{\rmii}[1]{{\mbox{\tiny\rm{#1}}}}

\newcommand{\im}{\mathop{\mbox{Im}}}

\newcommand{\Tint}[1]{{\hbox{$\sum$}\!\!\!\!\!\!\!\int\,}_{\!\!\!\!\raise-0.9ex\hbox{$\scriptstyle{#1}$}}}
\newcommand{\Tinti}[1]{{{\Sigma}\!\!\!\!\raise0.3ex\hbox{$\int$}_\rmii{${#1}$}}}

\newcommand{\bi}{\begin{itemize}}
\newcommand{\ei}{\end{itemize}}

\newcommand{\hide}[1]{ }
\newcommand{\hsl}[1]{\,\,\slash\!\!\!\!\!{#1}\,}
\newcommand{\bsl}[1]{\,\slash\!\!\!\!{#1}\,}
\newcommand{\msl}[1]{\,\slash\!\!\!{#1}\,}
\newcommand{\XPT}{$\chi$PT}
\def\TAsc(#1,#2)(#3,#4,#5)%
{\SetWidth{2.0}\CArc(#1,#2)(#3,#4,#5)\SetWidth{1.0}}
\def\Lwidth{3}

\def\TAgl(#1,#2)(#3,#4,#5){\SetWidth{2.0}\PhotonArc(#1,#2)(#3,#4,#5){\Lwidth}%
{6.283 #3 mul 360 div #4 #5 sub #4 #5 sub mul sqrt mul Tdensity mul}%
\SetWidth{1.0}}
\def\TLgl(#1,#2)(#3,#4){\SetWidth{2.0}\Photon(#1,#2)(#3,#4){\Lwidth}
{#1 #3 sub #1 #3 sub mul #2 #4 sub #2 #4 sub mul add sqrt Tdensity mul}%
\SetWidth{1.0}}
\newcommand{\piC}[1]{\;\parbox[c]{40pt}{\begin{picture}(120,60)(0,-20)
\SetWidth{1.0}\SetScale{0.35} #1 \end{picture}}\;}
\def\ConnectedA(#1,#2,#3){\piC{#1(60,-15)(75,34,146) #2(60,75)(75,214,326)%
 #3(60,60)(20,190,350)%
 \GBoxc(0,30)(10,10){1} \GBoxc(120,30)(10,10){1}%
  }}
\def\ConnectedB(#1,#2,#3){\piC{#1(60,-15)(75,34,146) #2(60,75)(75,214,326)%
 #3(60,60)(60,0)%
 \GBoxc(0,30)(10,10){1} \GBoxc(120,30)(10,10){1}%
  }}
\def\ConnectedC(#1,#2){\piC{#1(60,-15)(75,34,146) #2(60,75)(75,214,326)%
 \GBoxc(0,30)(10,10){1} \GBoxc(120,30)(10,10){1}%
  }}
\def\ConnectedD(#1,#2){\piC{#1(60,-15)(75,34,146) #2(60,75)(75,214,326)%
 \GBoxc(0,30)(10,10){1} \GBoxc(120,30)(10,10){1}%
 \SetWidth{2.0} 
 \Line(55,55)(65,65)%
 \Line(55,65)(65,55)
  }}

\makeatletter \@addtoreset{equation}{section} \makeatother
\renewcommand{\theequation}{\arabic{section}.\arabic{equation}}
\makeatletter
\renewcommand\section{\@startsection {section}{1}{\z@}%
                                   {-5.5ex \@plus -1ex \@minus -.2ex}
                                   {2.3ex \@plus.2ex}%
                                   {\normalfont\large\bfseries}}
\renewcommand\subsection{\@startsection{subsection}{2}{\z@}%
                                     {-3.25ex\@plus -1ex \@minus -.2ex}%
                                     {1.5ex \@plus .2ex}%
                                     {\normalfont\normalsize\bfseries}}
\renewcommand\thesection {\@arabic\c@section}
\renewcommand\thesubsection   {\thesection.\@arabic\c@subsection}
\renewcommand{\@seccntformat}[1]{%
\csname the#1\endcsname.\hspace{1.0em}}
\makeatother

\begin{document}

\begin{titlepage}
\begin{flushright}
BI-TP 2011/04\\
arXiv:1103.0372\\ 
\vspace*{1cm}
\end{flushright}
\begin{centering}
\vfill

{\Large{\bf
Heavy flavour kinetic equilibration in the confined phase
}} 

\vspace{0.8cm}

M.~Laine 

\vspace{0.8cm}

{\em
Faculty of Physics, University of Bielefeld, 
D-33501 Bielefeld, Germany\\}

\vspace*{0.8cm}

\mbox{\bf Abstract}
 
\end{centering}

\vspace*{0.3cm}
 
\noindent
By making use of a non-perturbative definition of 
a momentum diffusion coefficient as well as Heavy Meson Chiral 
Perturbation Theory, we investigate the Brownian motion
and kinetic equilibration of heavy quark flavours deep in the 
confined phase. It appears that the momentum diffusion coefficient 
can be expressed in terms of known low-energy constants; 
it increases rapidly at temperatures above 50 MeV, behaving as 
$\sim T^7/F_\pi^4$ for $\frac{m_\pi}{\pi} \ll T \ll F_\pi$, 
where $m_\pi$ and $F_\pi$ are the pion mass and decay constant, 
respectively. The early increase may suggest a broad peak in $\kappa/T^3$ 
around the QCD crossover. For a more detailed understanding 
the computation could be generalized in a number of ways.   

\vfill

 
\vspace*{1cm}
  
\noindent
April 2011

\vfill

\end{titlepage}

%
\section{Introduction}

An ideal scenario for what could happen in a heavy 
ion collision is that light quarks and gluons 
form a rapidly expanding thermalized medium, and that there 
are some ``probes'' available, whose properties are 
affected by the medium in a significant yet tractable way. Among 
the most attractive probe candidates are heavy quarks 
(charm and bottom quarks as well
as their antiparticles), which can be copiously produced in 
an initial hard process. After a while the heavy quarks decay, 
but modifications on their behaviour caused by the thermal 
medium could conceivably be deduced from the experimentally 
observed transverse momentum distributions and azimuthal 
anisotropies of the leptonic decay products~\cite{exp1,exp2}. 

With such motivations in mind, a significant body of work has 
been carried out during the last 20 years or so, concerning 
the effects that a thermal medium can have 
on the propagation of heavy quarks~\cite{bs,bt}. Roughly, 
the initial stages are characterized by radiative 
energy loss (bremsstrahlung), which may slow down the heavy quarks; 
the final stages are characterized by elastic scatterings, which
cause collisional energy loss but also produce
random kicks corresponding to Brownian motion. Assuming, 
idealistically, that the thermal system is spatially large 
enough for all of these processes to take place, they have been 
described by a number of related physical concepts 
and observables, such as energy loss (${\rm d}E/{\rm d}x$), 
stopping distance, jet quenching, (momentum) diffusion, drag, 
or kinetic equilibration. 

In the present paper, we focus on a late stage of the above 
scenario, in which the heavy quarks are practically at rest with 
respect to the thermal medium but also undergo Brownian motion.
(Possible extensions to more general kinematics are outlined
in the conclusions.) More specifically, we are interested
in an observable called the momentum diffusion coefficient, $\kappa$,
which characterizes the random force acting on the heavy 
quarks and, through a fluctuation-dissipation relation, also 
determines their kinetic equilibration rate. 

Recently, a number of theoretical works have addressed 
the same problem. For instance, $\kappa$ has been computed
to leading~\cite{tm} and next-to-leading~\cite{chm} order 
in the weak-coupling expansion. It has also been given  
a non-perturbative definition~\cite{eucl} in the framework 
of Heavy Quark Effective Theory~\cite{hqet1}--\cite{hqet2};  
this was partly inspired by computations through AdS/CFT 
techniques in the large-$\Nc$ limit of strongly 
coupled $\mathcal{N}=4$ Super-Yang-Mills theory, which suggest 
a larger $\kappa$ than in leading-order QCD~\cite{ads,ads2,ct}.
(Similar conclusions have also been reached for AdS models 
resembling QCD; see e.g.\ ref.~\cite{ek} and references therein.) 
Numerical simulations have 
been carried out within so-called classical lattice gauge theory, 
suggesting again a value larger than indicated by the weak-coupling
expansion~\cite{mink}. 
All of this makes a strong case for attacking the problem with 
lattice simulations, a challenge that may be less daunting that 
the determination of many other ``transport coefficients'', such as 
viscosities, because of the simpler structure of the pertinent
spectral function~\cite{rhoE}; indeed the very first numerical 
attempts look rather promising~\cite{hbm2}. The inherent 
uncertainties related to analytic continuation from Euclidean 
numerical data have also been looked into, and it appears that 
it might be feasible to carry through the program 
at least on the qualitative level~\cite{analytic}.
Once a value of $\kappa$ is available, it can be incorporated
in hydrodynamical simulations to yield results 
relevant for experiment (cf.\ e.g.\ ref.~\cite{tm}); there 
is a growing body of such works under way. 

The purpose of the present paper is to make use of the
non-perturbative definition of $\kappa$ introduced in 
ref.~\cite{eucl}, and to evaluate it deep in the confined phase. 
This is possible in QCD because, due to chiral symmetry 
breaking,\footnote{%
 The number of light flavours is assumed non-zero, $\Nf > 0$; 
 it is unclear whether $\kappa$ can be given a sensible
 meaning in the confined phase for $\Nf = 0$~\cite{eucl} 
 even if the related Euclidean correlator appears
 to exist~\cite{rhoE}.
 } 
the infrared dynamics of the confined phase
can be parametrized with a small number
of ``low-energy constants''. In fact, in a certain limit, 
we find that $\kappa$ is fully 
determined in terms of the pion decay constant and the 
pion mass. There is, of course, a long history to applying 
Chiral Perturbation Theory (not to mention hadronic models)
for the computation of various thermodynamic properties of 
the pion gas, see e.g.\ ref.~\cite{gele}; 
past developments and the non-trivial theoretical 
challenges that are related  particularly to transport coefficients have 
recently been discussed in the context of the bulk 
viscosity of strongly interacting matter in ref.~\cite{lgdm}. 

The paper is organized as follows. 
The non-perturbative definition of the momentum diffusion 
coefficient is reviewed in \se\ref{se:basic}; the chiral 
effective theory relevant for handling heavy-light mesons
in the confined phase is described in \se\ref{se:hmxpt}; 
and the computation of the momentum diffusion coefficient
is presented in \se\ref{se:comp}. Some conclusions
comprise \se\ref{se:concl}.

%
\section{Basic physics of momentum diffusion}
\la{se:basic}

Considering time scales short compared with the life-time of 
the heavy quarks (or heavy-light mesons)
but long compared with those of microscopic
processes involving gluons and light quarks (or light mesons), 
which have a momentum $p\sim T \sim 100$~MeV, where $T$ 
denotes the temperature, the dynamics of the heavy
degrees of freedom, with a mass $M \gg T$, presumably resembles 
Brownian motion. If so, it can be described by the Langevin 
equation, with the role of the stochastic noise being 
played by a force induced by QCD-mediated collisions 
with gluons and light quarks (or light mesons), 
which are present with an abundant number density $n\sim T^3$. 
The corresponding equations of motion have the form 
\ba
 \dot{p}_k(t) 
 &  = &   
 -\eta_\rmii{D} \, p_k(t) + \xi_k(t) 
 \;, \la{Lan1} \\ 
 \langle\!\langle \, \xi_k(t) \, \xi_l(t') \, \rangle\!\rangle & = & 
 \kappa \, \delta_{kl} \, \delta(t-t')
 \;, \qquad
 \langle\!\langle \xi_k(t) \rangle\!\rangle = 0 
 \;, \la{Lan2}
\ea
where $p_k$ is the momentum of the heavy objects ($k=1,2,3$);
$\xi_k$ is a Gaussian stochastic noise; and 
$\langle\!\langle ... \rangle\!\rangle$ denotes 
an average over the noise. According to \eq\nr{Lan2} 
the momentum diffusion 
coefficient, $\kappa$, characterizes 
the auto-correlator of the force, 
\be
 \kappa = \fr13  \int_{-\infty}^{\infty} \! {\rm d}t \, 
 \sum_k \langle\!\langle \, \xi_k(t) \, \xi_k(0) \, \rangle\!\rangle
 \;, \la{kappa_cl}
\ee 
whereas the coefficient $\eta_\rmii{D}$
appearing in \eq\nr{Lan1} is referred to as the ``kinetic 
equilibration rate'' or the ``drag coefficient''. 
As is well-known, in classical statistical physics the two 
can be fluctuation-dissipation related to each other:
$\eta_\rmii{D} \simeq \kappa / (2\, T M)$, 
where we interpret $M$ as a specific mass definition 
(more precisely the heavy quark ``kinetic'' mass; 
we assume a regularization scheme respecting Lorentz symmetry, 
and then $M$ is equal to the ``rest'' or ``pole'' mass; 
it should be fixed non-perturbatively). 

Now, it was argued in ref.~\cite{eucl} that the classical 
definition of $\kappa$ as an autocorrelator of forces, \eq\nr{kappa_cl}, 
can ``naturally'' be extended to QCD. Suppose that  
we know the heavy quark Hamiltonian, 
$\hat H$, as well as the Noether current associated with the U(1) 
flavour symmetry, $\hat{\mathcal{J}}^\mu$. More specifically, 
$\hat H$ and $\hat{\mathcal{J}}^0$ are needed up to $\rmO(M^0)$
in an expansion in a large $M$, 
whereas the $\hat{\mathcal{J}}^k$ are needed up to $\rmO(1/M)$.
With these operators we can define a ``susceptibility''
related to the conserved charge, 
\be
 \chi^{00} \equiv
 \beta \int_{\vec{x}} \,
 \left\langle
  \hat {\cal{J}}^0(t,\vec{x}) \,  
  \hat {\cal{J}}^0(t,\vec{0})
 \right\rangle_T
 \;,
  \quad  \beta \equiv \frac{1}{T}
 \;,
  \quad \int_{\vec{x}} \equiv \int \! {\rm d}^{3} \vec{x}
 \;, \la{susc}
\ee
and the ``acceleration'' associated with the spatial components, 
\be
 \frac{{\rm d}\hat{\cal J}^k}{{\rm d}t}
 = i \bigl[\hat H, \hat{\cal J}^k \bigr] + 
 \frac{\partial\hat{\cal J}^k}{\partial t}
 \;, \la{eom}
\ee
where the partial derivative acts on possible background fields. 
Consequently, 
\be
 \kappa \;\equiv\; 
 \frac{\beta}{3}  \sum_{k=1}^{3} 
 \lim_{\omega\to 0}
  \biggl[
  \lim_{M\to\infty} \frac{M^2}{\chi^{00}}
  \int_{-\infty}^\infty 
 \!\! {\rm d}t \,
  e^{i \omega (t-t')} \!
 \int_{\vec{x}} \,
 \biggl\langle
  \fr12 \biggl\{ 
  \frac{{\rm d} \hat {\cal{J}}^k(t,\vec{x}) }{{\rm d}t}, 
  \frac{{\rm d} \hat {\cal{J}}^k(t',\vec{0})}{{\rm d}t'}
   \biggr\} 
 \biggr\rangle_T
 \biggr]
 \;. \la{kappa_def}
\ee
Since $M  \int_\vec{x} {\rm d} \hat{\mathcal{J}}^k / {\rm d} t$ 
represents, according to Newton's law, a force, this definition is 
indeed a generalization of \eq\nr{kappa_cl}. The ordering of the 
various limits requires, however, a careful analysis~\cite{eucl}; 
the situation simplifies only if a dependence 
$
 {\rm d} \hat {\cal{J}}^k / {\rm d}t \sim 1/M 
$
can be factored out and cancelled. 
We also note that, as dictated by standard relations 
between various time orderings at finite temperatures~\cite{ftft},  
\be
 \kappa = \lim_{\omega\to 0} \frac{2 T \rho_\rmii{E}(\omega)}{\omega}
 \;, \la{relation}
\ee
where the spectral function $\rho_\rmii{E}$
is defined as in \eq\nr{kappa_def}
but with a commutator replacing the anticommutator. 
(The notation $\langle \ldots \rangle^{ }_T$ 
refers to the usual thermal average.)

%
\section{Heavy Meson Chiral Perturbation Theory}
\la{se:hmxpt}

The goal now would be to evaluate \eq\nr{kappa_def} 
at low temperatures. This can be achieved by making use of an effective
field theory that is valid in the regime considered and allows us
to define the operators entering the definition. Given that 
the framework may be unfamiliar within the 
finite-temperature community, we give a few ingredients in 
the following, although no attempt is made at a comprehensive 
review (see refs.~\cite{mw_rev,Casalbuoni,book} for introductions). 

The ``usual'' chiral Lagrangian describing the pseudo Nambu-Goldstone
bosons of chiral symmetry breaking has the form~\cite{gl} 
\be
 \mathcal{L}_\rmi{\XPT} = 
 \frac{F^2}{4} \tr (\partial^\mu U \partial_\mu U^\dagger)  
 + \frac{\Sigma}{2} \, \tr(\mathcal{M}^\dagger U + U^\dagger \mathcal{M})
 + \ldots
 \;, \la{L_XPT}
\ee
where $F$ is the pion decay constant 
in the chiral limit (naively $F\simeq 93$~MeV), 
$\Sigma$ is the chiral condensate, $\mathcal{M}$ is the quark 
mass matrix, and $U \in \mbox{SU}(\Nf)$ is the Goldstone field. 
For simplicity we take the mass matrix to be of the 
form $\mathcal{M} = m\, \mathbbm{1}$ in the following, 
with $m\in \mathbbm{R}$; then the pion mass squared is
$m_\pi^2 = 2 m \Sigma/F^2 + \rmO(1/F^4)$. 
Parametrizing 
$
 U = \exp (\frac{2i\xi}{F}) 
$, 
the tree-level propagator is 
\be 
 \langle 
   \,\xi^{ }_{ab}(x)\, \xi^{ }_{cd}(y)\,  
 \rangle^{(0)} = 
 \fr12 
 \Bigl( 
   \delta^{ }_{ad} \delta^{ }_{bc} -
   \frac{1}{\Nf} \delta^{ }_{ab} \delta^{ }_{cd}
 \Bigr) \Delta(x-y;m_\pi^2)
 \;, \la{G_prop}
\ee
where $\Delta$ is a scalar propagator.
It is important to keep in mind that for the 
$\mathcal{M}$ chosen, \eq\nr{L_XPT} contains no terms 
cubic in $\xi$, i.e.\ no three-pion interactions.
(Interactions among odd numbers of pions
are contained in the Wess-Zumino-Witten term, 
but it is suppressed by a large number of derivatives.) 

The next step is to supplement the chiral Lagrangian with 
a piece describing heavy-light mesons. The resulting effective
description is called Heavy Meson Chiral Perturbation Theory 
(HM$\chi$PT)~\cite{bdo,hmxpt1,hmxpt2}. 

To start with we note that, unlike 
Heavy Quark Effective Theory (HQET)~\cite{hqet1}--\cite{hqet2}, 
which describes 
all states containing a single heavy quark or antiquark 
(provided that the associated gluons and light quarks are soft),
the mesonic HM$\chi$PT\ only contains a specific subset of such 
states, namely parity-odd pseudoscalar and vector mesons ($D$ and $D^*$ 
for charm, ${B}$ and ${B}^*$ for bottom), with total spin 0 or 1. 
The standard convention is to denote the scalar by $P_a$ 
and the vector by $P^{*\mu}_a$, where $a$ is a light flavour
index ($a = 1,\ldots,\Nf$); to streamline the notation we 
define $Q^\mu_a \equiv P^{*\mu}_a$ in the following. 
Although $Q^{\mu}_a$ is written as a four-vector, it only
contains three independent components, cf.\ \eq\nr{perp} below.

The form of HM\XPT\ is dictated by symmetries, of which there 
are many. First of all there is a U(1) symmetry related to
the conserved heavy quark number; this yields the Noether current
$\mathcal{J}^\mu$ alluded to above. Second, HQET at $\rmO(M^0)$
displays a heavy quark spin symmetry, since Pauli matrices 
first appear at $\rmO(1/M)$; this implies a relation between 
the fields $P_a$ and $Q^{\mu}_a$ (cf.\ \eq\nr{symm} below). 
Third, the light-quark index $a$ 
enjoys a specific transformation property  under
the SU($\Nf$)$_L \times$SU($\Nf$)$_R$ chiral symmetry. 
Fourth, even though the effective Lagrangian is non-relativistic, 
its origin in relativistic QCD implies that it also remembers 
something about proper Lorentz symmetry, provided that this has
not been broken through regularization. This is usually implemented
by introducing an arbitrary parameter, a four-velocity $v^\mu$, 
and an associated ``velocity reparametrization invariance''~\cite{lm}.
Finally, there are the usual discrete C, P, and T symmetries of QCD. 

It turns out to be non-trivial to implement all the symmetries 
in a convenient way. 
In fact, in order to define a simple parity transformation it is 
not practical to use the Goldstone field $U$ in the part of 
the Lagrangian involving the heavy mesons, but rather
a ``coset field'' $\sqrt{U}$.\footnote{%
  We follow the notation of e.g.~ref.~\cite{bh}, where 
  HM\XPT\ was used in connection 
  with a lattice investigation. 
 }
(General formal considerations can be 
found in refs.~\cite{old1,old2}.)
If $U$ transforms in SU($\Nf$)$_L \times$SU($\Nf$)$_R$ as
\be
 U \to L\, U R^\dagger
 \;, 
\ee
then the field $\sqrt{U}$ can be assigned the transformation
\be
 \sqrt{U} \to L \sqrt{U} W^\dagger 
  \quad \mbox{and} 
  \quad
 \sqrt{U} \to W \sqrt{U} R^\dagger  
 \;, \la{new_trafo}
\ee
where the complicated (space-time dependent) field $W$ is defined
through \eqs\nr{new_trafo}. We can subsequently introduce
the traceless and Hermitean fields 
\ba
 \mathcal{V}_\mu & \equiv & 
  \frac{i}{2}
  \Bigl[
    \sqrt{U}^\dagger \partial_\mu \sqrt{U}
  + \sqrt{U} \partial_\mu \sqrt{U}^\dagger
  \Bigr] 
 \;, \la{def_VA_1}\\ 
  \mathcal{A}_\mu & \equiv & 
  \frac{i}{2}
  \Bigl[
    \sqrt{U}^\dagger \partial_\mu \sqrt{U}
  - \sqrt{U} \partial_\mu \sqrt{U}^\dagger
  \Bigr]
 \;, \la{def_VA_2}
\ea
which transform as 
\be
 \mathcal{V}_\mu \to W \mathcal{V}_\mu W^\dagger +
                   i\, W \partial_\mu W^\dagger 
 \;, \quad
 \mathcal{A}_\mu \to W \mathcal{A}_\mu W^\dagger
 \;. \la{trafo_VA}
\ee
Note that in the chiral expansion, inserting 
$ 
  \sqrt{U} = \exp (\frac{i\xi}{F}) 
$, 
we get
\be
 \mathcal{V}_\mu = \frac{i}{2F^2}
 \Bigl( \xi\, \partial_\mu \xi - \partial_\mu \xi\, \xi \Bigr) 
   + \rmO\Bigl( \frac{\xi^4}{F^4} \Bigr)
 \;, \quad
 \mathcal{A}_\mu = - \frac{1}{F}\, \partial_\mu \xi
   + \rmO\Bigl( \frac{\xi^3}{F^3} \Bigr) 
 \;, \la{lo_VA}
\ee
i.e.\ 
$\mathcal{V}_\mu$ couples to an even number of pions and 
$\mathcal{A}_\mu$ to an odd number. 

The heavy meson fields $P_a$ and $Q^{\mu}_a$ are normally
assembled into a $4\times 4$ matrix $H_a$ and its conjugate
$\widebar{H}_a \equiv \gamma^0 H_a^\dagger \gamma^0$:
\be
 H_a \equiv \frac{1 + \msl{v}}{2} 
 \Bigl( \bsl{Q}_{\! a} + i \gamma_5 P_a \Bigr)
 \;, \quad
 \widebar{H}_a = 
 \Bigl( \bsl{Q}^\dagger_{\! a} + i \gamma_5 P_a^\dagger \Bigr)
 \frac{1 + \msl{v}}{2}
 \;, \la{Ha_def}
\ee
where, following a frequent convention, 
$(...)^\dagger$ denotes complex conjugation 
acting on $P_a$, $Q^\mu_a$.
Here $v^\mu$ is a four-velocity with $v\cdot v = 1$.
The vector meson field is constrained by
\be
 v\cdot Q_a = 0
 \;, \la{perp}
\ee
which implies for instance that 
$
 \msl{v} \bsl{Q}_{\! a} =  - \, \bsl{Q}_{\! a} \! \msl{v}   
$.
The chiral transformation can be assigned as 
\be
 H \to H \, W^\dagger
 \;, \quad
 \widebar{H} \to W\, \widebar{H}
 \;.
 \la{trafo_H}
\ee
Defining 
\be
 \mathcal{D}^\mu_{ba} 
  \equiv
 \partial^\mu\, \delta^{ }_{ba} + i \mathcal{V}^\mu_{ba}
 \;, \la{def_Dmu}
\ee
the effective Lagrangian at $\rmO(M^0)$ has the form
\ba
 \mathcal{L}_\rmi{HM\XPT}^{(0)} & = &  
 - i \, \tr (\widebar{H}_a \, v\cdot\mathcal{D}^{ }_{ba} \, H^{ }_b)
 + M^{ }_P \, \tr (\widebar{H}_a H_a) 
 + g_\pi \, \tr (\widebar{H}_a H^{ }_b \, 
     \hsl{\mathcal{A}}^{ }_{\! ba} \gamma_5)
 \nn & & +\, 
 \sigma_1 \, \tr (\widebar{H}_a H_b)
 (\sqrt{U} \mathcal{M}^\dagger \sqrt{U} + 
  \sqrt{U}^\dagger \mathcal{M} \sqrt{U}^\dagger)_{ba} 
 \nn & & +\, 
  \sigma_1' \, \tr (\widebar{H}_a H_a)
 (\sqrt{U} \mathcal{M}^\dagger \sqrt{U} + 
  \sqrt{U}^\dagger \mathcal{M} \sqrt{U}^\dagger)_{bb} 
 + \ldots
 \;, \la{L_0}
\ea
where the trace is over Dirac matrices, and ``$\ldots$'' 
denotes terms of higher order in the chiral expansion. 
The coefficient $g_\pi$ is a new dimensionless low-energy 
constant, estimated from e.g.\ $D^*\to D\pi$ decays~\cite{book} or 
lattice determinations (e.g.\ refs.~\cite{lat1,lat2}) 
to be around $g_\pi \simeq 0.5$.

The mass term on the first row of \eq\nr{L_0}, 
proportional to $M^{ }_P$, is often not shown, because
it can be shifted away by a time-dependent phase transformation. 
However, for the finite-temperature application that will be 
discussed in the next section, it is useful to keep $M^{ }_P$ explicit. 
The terms proportional to the low-energy constants $\sigma_1, \sigma_1'$
play a role, at leading order, for the mass spectrum: $\sigma_1'$
shifts $M^{ }_P$ by a flavour-independent amount, whereas $\sigma_1$ would 
break the flavour symmetry if $\mathcal{M}$ were non-degenerate. 
In our case both of these effects can be accounted for through $M^{ }_P$, 
so we omit $\sigma_1, \sigma_1'$ in the following. The spin symmetry 
between $P_a$ and $Q^\mu_a$ is only broken at $\rmO(1/M)$; although 
we presently turn to such effects, the ``trivial'' mass splitting that  
they induce will be considered to already be accounted for 
by the present remarks, and we denote the vector meson mass by $M^{ }_Q$. 

Now, we turn to terms of $\rmO(1/M)$.
Some of these can be deduced from \eq\nr{L_0} through 
reparametrization invariance~\cite{lm}. This is a transformation
which leaves invariant the ``original'' momentum operator
$
 P_\mu \equiv M v_\mu + i \mathcal{D}_\mu
$
as well as scalar products such as $v\cdot v = 1$. Writing 
$
 v_\mu \to v_\mu + \epsilon_\mu / M
$
with $v\cdot \epsilon = 0$,  
which implies
$
 i \mathcal{D}_\mu \to i \mathcal{D}_\mu  - \epsilon_\mu
$, 
we observe in particular that the combination
\be
 \mathcal{O}_{ba} = 
 i v\cdot \mathcal{D}_{ba} 
 - \frac{a_1}{2 M}\, (\mathcal{D}\cdot\mathcal{D})_{ba}  
 \la{rel_str}
\ee
is invariant only for $a_1 = 1$. 
This is clearly a reflection of Lorentz symmetry, and means that 
the coefficient of a particular $\rmO(1/M)$-operator is fixed
in terms of that of $i v\cdot \mathcal{D}_{ba}$ at $\rmO(M^0)$. 
(Whether it is $M$ or $M^{ }_P$ that appears in the denominator
of \eq\nr{rel_str} plays no role, because their difference 
amounts to an effect of $\rmO(1/M^2)$.) So, 
\be
 \mathcal{L}_\rmi{HM\XPT}^{(1)} = 
 \frac{1}{2M} 
 \, \tr [ \widebar{H}^{ }_a \, 
 (\mathcal{D}\cdot\mathcal{D})^{ }_{ba} \, H^{ }_b ]
 + \ldots
 \;. \la{L_1}
\ee
The term proportional to $g_\pi$ in \eq\nr{L_0} also 
implies the existence of terms of $\rmO(g_\pi/M)$ with specific 
coefficients, but the argument is not quite as transparent
as that with \eq\nr{rel_str}; 
we postpone a discussion to after \eq\nr{symm}.
Analyses of all terms of $\rmO(1/M)$, also those not fixed by
reparametrization invariance, have been carried out
in refs.~\cite{invM,bg}.

%
\section{Main computation}
\la{se:comp}

We now apply the theory defined by \eqs\nr{L_0}, \nr{L_1} to estimate 
the momentum diffusion coefficient as defined by \eq\nr{kappa_def}. 
The computation carried out is rather straightforward and simple-minded, 
but the result is interesting so we present the steps in some detail. 

%
\subsection{Force operator}

We start by rewriting the sum of \eqs\nr{L_0}, \nr{L_1} after 
carrying out the Dirac traces, setting $v \to (1,\vec{0})$, and 
adding terms of $\rmO(g_\pi/M)$ through an argument 
to be discussed presently
(the indices $k,l,m$ are spatial; indices are
raised and lowered with the metric ($+$$-$$-$$-$); 
repeated indices are summed over; and
``$\ldots$'' indicates terms of higher order in the chiral expansion):
\ba
 \mathcal{L}_\rmi{HM\XPT} & = & 
 2 \Bigl[ 
   P_a^\dagger \, i \mathcal{D}^0_{ba} P^{ }_b 
  -  
   M^{ }_P \,    P_a^\dagger P^{ }_a 
 \Bigr]
 + 2 \Bigl[ 
   Q^\dagger_{ak} \, i \mathcal{D}^0_{ba} Q^{ }_{bk} 
 -
   M^{ }_Q \,    Q_{ak}^\dagger Q^{ }_{ak}
 \Bigr]
 \nn 
 & + & 
 2 i  g_\pi \, \Bigl[ 
   P_a^\dagger Q^{ }_{bl}
  - Q^\dagger_{al} P^{ }_b
    + \epsilon^{ }_{klm}
   Q^\dagger_{ak} Q^{ }_{bm}
 \Bigr] \, \mathcal{A}^l_{ba}
 \nn 
 & + & 
 \frac{1}{M}
 \Bigl[
   P_a^\dagger (\mathcal{D}^l  \mathcal{D}^l)^{ }_{ba} P^{ }_b 
  +  
   Q^\dagger_{ak} (\mathcal{D}^l \mathcal{D}^l)^{ }_{ba} Q^{ }_{bk} 
   \Bigr] 
 \nonumber \\[1mm] 
 & + & 
 \frac{g_\pi}{M}
  \Bigl[
    2 P^\dagger_a \mathcal{D}^k_{cb} Q^{ }_{ck}
  + 2 Q^\dagger_{ck} \overleftarrow{\mathcal{D}}^k_{ac} P_b
  + \epsilon^{ }_{klm}
  \Bigl(   Q^\dagger_{ak} \mathcal{D}^l_{cb} Q_{cm} 
  -  Q^\dagger_{ck} \overleftarrow{\mathcal{D}}^l_{ac} Q_{bm}  
  \Bigr)
  \Bigr] \mathcal{A}^0_{ba}
 \nn 
 & + & \rmO\Bigl( \frac{1}{M} \Bigr)  + \ldots
 \;. \la{L_expl}
\ea
In the part of $\rmO(1/M)$ only terms with 
spatial derivatives acting on $P$ and $Q$  have been shown, 
because these are the only ones contributing to $\mathcal{J}^k$.
Also, 
$
 Q^\dagger_{c} \overleftarrow{\mathcal{D}}^k_{ac}
 \equiv
 ( \mathcal{D}^k_{ca} Q^{ }_c )^\dagger
$. 

The heavy quark spin symmetry amounts to the transformations
\be
 \delta P^{ }_a = -\alpha^{ }_l\, Q^{ }_{al}
 \;, \quad
 \delta Q^{ }_{ak} = \alpha^{ }_k\, P^{ }_a + 
 \epsilon^{ }_{klm} \alpha^{ }_l\, Q^{ }_{am}
 \;, \quad
 \alpha^{ }_k \in \mathbbm{R} 
 \;, \quad
 k,l,m \in \{ 1,2,3 \}
 \;. \la{symm}
\ee
Terms of $\rmO(M^0)$ in \eq\nr{L_expl}
are easily seen to be invariant under this transformation; 
the mass terms are invariant only if $M^{ }_Q = M^{ }_P$, i.e.\ 
the difference $M^{ }_Q - M^{ }_P$ must be of $\rmO(1/M)$, as 
mentioned above. Even though this is not the case 
in general, the terms of $\rmO(1/M)$ shown
explicitly in \eq\nr{L_expl}
are also invariant under \eq\nr{symm}
(the term on the fourth row is invariant
only up to total derivatives and 
terms containing covariant derivatives acting on $\mathcal{A}^0_{ba}$, 
which have been omitted because they do not contribute
to $\mathcal{J}^k$). The invariance exists because these
terms are related to invariant terms of $\rmO(M^0)$.

The fourth row of \eq\nr{L_expl}, with a coefficient 
$g_\pi/M$, is new with respect to \eq\nr{L_1}.  It is connected to 
the second row  of \eq\nr{L_expl}
through reparametrization invariance; to be concrete, 
one could imagine 
that the vector field of an original relativistic 
theory, let us denote it with $\tilde{Q}$, satisfy 
a transversality constraint of the type $P_\mu \tilde{Q}^\mu = 0$ 
rather than \eq\nr{perp}. So, 
recalling that 
$
  P_\mu \equiv M v_\mu + i \mathcal{D}_\mu
$, 
every appearance of $Q^\mu$ can be 
replaced with $\tilde{Q}^\mu \equiv Q^\mu  - v^\mu i \mathcal{D} \cdot Q/M$, 
where $Q^\mu$ does satisfy \eq\nr{perp}~\cite{lm}. 
This yields terms of $\rmO(g_\pi/M)$ with fixed coefficients.
The $\rmO(1/M)$ structures shown in \eq\nr{L_expl} that have been 
determined through reparametrization invariance are the only 
ones from the full list~\cite{bg} that 
contain spatial derivatives acting on $P_a,Q_{al}$, 
and therefore contribute to $\mathcal{J}^k$. 
(An explicit crosscheck on the logic presented 
could be obtained by starting from a genuinely 
relativistic formulation, such as the ones 
in refs.~\cite{invM,bj}, 
and taking the non-relativistic limit only in the end, 
because then all consequences of Lorentz symmetry are 
automatically respected. In the following the terms 
of $\rmO(g_\pi/M)$
are included for illustration but are omitted from the final result.) 

Now, all terms proportional to $g_\pi$ 
couple to $\mathcal{A}^\mu_{ba}$ and thus to an odd number of pions, 
whereas terms without $g_\pi$ couple to an even number of pions 
(cf.\ \eq\nr{lo_VA}). Also, $g_\pi$ necessarily mixes the
fields $P^{ }$ and $Q_{ }^\dagger$, 
or  $Q^{ }$ and $Q_{ }^\dagger$. 
This makes the terms proportional to $g_\pi$ in general 
the most important ones from the point of view of zero-temperature 
phenomenology, such as the decays of the $D^*$ and $B^*$ mesons~\cite{book};
a relevant amplitude is illustrated in \fig\ref{fig:graphs}(a).
However, if all particles are on-shell, this process is  
kinematically forbidden for $B$-mesons, because the mass difference
between $B$ and $B^*$, scaling as $1/M$, is below the pion mass. 
Therefore the dominant processes for $B$-mesons are of the types shown 
in \figs\ref{fig:graphs}(b) and \ref{fig:graphs}(c), and 
the scattering amplitude is $\rmO(1/F^2)$.

%
\begin{figure}[t]

\hspace*{0.4cm}
\begin{minipage}[c]{3.5cm}
\begin{picture}(100,80)(-50,-40)
\SetScale{0.6}  
\SetWidth{2.0}
\Line(-40,-30)(0,0)%
\DashLine(-40,30)(0,0){6}%
\Line(-2,-2)(40,-2)%
\Line(-2,2)(40,2)%
\Text(-32,-21)[c]{$P$}%
\Text(-32,22)[c]{$\pi$}%
\Text(32,0)[c]{$P^*$}%
\Text(0,18)[c]{$\displaystyle\frac{g_\pi}{F}$}%
\Text(0,-40)[c]{(a)}
\end{picture}
\end{minipage}%
\hspace*{0.5cm}%
\begin{minipage}[c]{5.0cm}
\begin{picture}(120,80)(-60,-40)
\SetScale{0.6}  
\SetWidth{2.0}
\Line(-60,-30)(0,0)%
\Line(60,-30)(0,0)%
\DashLine(-60,30)(0,0){6}%
\DashLine(60,30)(0,0){6}%
\Text(-47,-21)[c]{$P$}%
\Text(-47,22)[c]{$\pi$}%
\Text(47,-21)[c]{$P$}%
\Text(47,22)[c]{$\pi$}%
\Text(0,-15)[c]{$\displaystyle\frac{1}{F^2}$}%
\Text(0,-40)[c]{(b)}
\end{picture}
\end{minipage}%
\hspace*{0.5cm}%
\begin{minipage}[c]{5.0cm}
\begin{picture}(120,80)(-60,-40)
\SetScale{0.6}  
\SetWidth{2.0}
\Line(-60,-30)(-20,0)%
\Line(60,-30)(20,0)%
\Line(-22,2)(22,2)%
\Line(-22,-2)(22,-2)%
\DashLine(-60,30)(-20,0){6}%
\DashLine(60,30)(20,0){6}%
\Text(-47,-21)[c]{$P$}%
\Text(-47,22)[c]{$\pi$}%
\Text(47,-21)[c]{$P$}%
\Text(47,22)[c]{$\pi$}%
\Text(-12,16)[c]{$\displaystyle\frac{g_\pi}{F}$}%
\Text(12,16)[c]{$\displaystyle\frac{g_\pi}{F}$}%
\Text(0,-10)[c]{$P^*$}%
\Text(0,-40)[c]{(c)}
\end{picture}
\end{minipage}%

\vspace*{0.5cm}

\caption[a]{\small 
Some of the scatterings experienced by $P$-mesons 
almost at rest with respect to a pion gas
($P$ and $P^*$ could represent $B$ and $B^*$-mesons, respectively). 
The process (a) is kinematically allowed for $D$-mesons, 
but not for $B$-mesons. 
There is no $t$-channel scattering on thermal pions, 
because the Lagrangian in \eq\nr{L_XPT} contains no three-pion 
vertex and that in \eq\nr{L_expl} contains no $PP\pi$-vertex. 
Electromagnetic interactions, such as scatterings
on blackbody photons, have been omitted.  
} 
\la{fig:graphs}
\end{figure}
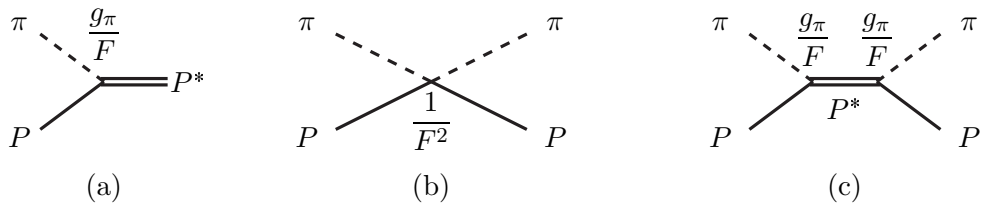
%

Next, we obtain from \eq\nr{L_expl} 
the U(1) Noether current: 
\ba
 \mathcal{J}^0 & = & 2 P^\dagger_a P^{ }_a 
 + (P^{ }_a \leftrightarrow Q^{ }_{ak} )
 + \rmO\Bigl( \frac{1}{M} \Bigr) 
 \;, \la{J_0} \\ 
 \mathcal{J}^k & = & \frac{1}{M} 
 \Bigl[
  i (P_a^\dagger \partial^k P^{ }_a - \partial^k P^\dagger_a \, P^{ }_a ) 
  - 2 P^\dagger_a \mathcal{V}^k_{ba} P^{ }_b
 \Bigr] + (P^{ }_a \leftrightarrow Q^{ }_{al} )
 \nn & + &  
 \frac{2ig_\pi}{M}
 \Bigl[P^\dagger_a Q^{ }_{bk} - Q^\dagger_{ak} P^{ }_b
 + \epsilon^{ }_{lkm} Q^\dagger_{al} Q^{ }_{bm}
    \Bigr] \mathcal{A}^0_{ba}
 + \rmO\Bigl( \frac{1 }{M^2} \Bigr) 
 \;. \la{J_k}
\ea
In order to determine the acceleration from \eq\nr{eom}, 
we would need the Hamiltonian; however, equivalent information
should be contained in classical equations of motion, which read 
\ba
 i \partial_0 P^{ }_a & = & M^{ }_P \, P_a + 
 \mathcal{V}^0_{ba} \, P^{ }_b
 - i g_\pi \mathcal{A}^l_{ba} Q^{ }_{bl}  
 + \rmO\Bigl( \frac{1}{M} \Bigr)
 \;, \\
 i \partial_0 Q^{ }_{ak} & = &
 M^{ }_Q \, Q^{ }_{ak} + 
 \mathcal{V}^0_{ba} \, Q^{ }_{bk}
 + i g_\pi \Bigl( 
    \mathcal{A}^k_{ba} P^{ }_{b}
  - \epsilon^{ }_{klm} \mathcal{A}^l_{ba} Q^{ }_{bm}  
 \Bigr)
 + \rmO\Bigl( \frac{1}{M} \Bigr)
 \;. \la{eom_2} 
\ea
Note that when acting on $\mathcal{J}^0$, $\mathcal{J}^k$, 
the terms of $\rmO(M^{ }_P, M^{ }_Q)$ from here cancel out
(apart from a possible remainder proportional to 
$M^{ }_Q - M^{ }_P \sim 1/M$), so that effectively 
the time derivatives count as terms of $\rmO(M^0)$.

Combining \eqs\nr{J_k}--\nr{eom_2}, and noting that 
in \eq\nr{kappa_def} a spatial average can be taken over
the currents and that therefore partial integrations are 
allowed with respect to $\partial^k$, we finally obtain 
$ \partial^0 \int_{\vec{x}} \mathcal{J}^k$. 
Terms do proliferate quite a bit: defining
\ba
 i F^{\mu\nu}_{ab} & \equiv &
  [\mathcal{D}^{\mu\,T},\mathcal{D}^{\nu\,T}]_{ab}
 = i \Bigl(
       \partial^\mu \mathcal{V}^\nu_{ba}
     - \partial^\nu \mathcal{V}^\mu_{ba}
     + i \mathcal{V}^\mu_{ca} \mathcal{V}^\nu_{bc}
     - i \mathcal{V}^\nu_{ca} \mathcal{V}^\mu_{bc}
     \Bigr)
 \;, \la{F_def} \\ 
 i G^{\mu\nu}_{ab} & \equiv &
  [\mathcal{D}^{\mu\,T},i \mathcal{A}^{\nu\,T}]_{ab}
 = i \Bigl(
       \partial^\mu \mathcal{A}^\nu_{ba}
     + i \mathcal{V}^\mu_{ca} \mathcal{A}^\nu_{bc}
     - i \mathcal{A}^\nu_{ca} \mathcal{V}^\mu_{bc}
     \Bigr)
 \;, \la{G_def} \\ 
 i H^{\mu\nu}_{ab} & \equiv & 
 [i \mathcal{A}^{\mu\,T},i \mathcal{A}^{\nu\,T}]_{ab}
 = i \Bigl(
      i \mathcal{A}^\mu_{ca} \mathcal{A}^\nu_{bc}
     - i \mathcal{A}^\nu_{ca} \mathcal{A}^\mu_{bc}
     \Bigr)
 \;, \la{H_def} \\ 
 i \tilde{H}^{\mu\nu}_{ab} & \equiv & 
 \{ i \mathcal{A}^{\mu\,T},i \mathcal{A}^{\nu\,T} \}_{ab}
 = i \Bigl(
      i \mathcal{A}^\mu_{ca} \mathcal{A}^\nu_{bc}
     + i \mathcal{A}^\nu_{ca} \mathcal{A}^\mu_{bc}
     \Bigr)
 \;, \la{tH_def}
\ea
where $(..)^T$ denotes a transpose with respect to flavour 
indices,  we obtain the structure 
\ba
 \partial^0 \int_{\vec{x}} \mathcal{J}^k & = &  
 \frac{2}{M} \int_{\vec{x}}
 \biggl\{ 
    P^\dagger    (F^{k0} - g_\pi^2 H^{k0}) P^{ }
  + Q_l^\dagger  (F^{k0} - g_\pi^2 H^{k0}) Q^{ }_l 
 \nn & + &  
 i g_\pi 
 \Bigl[ 
  P^\dagger \, G^{lk} \, Q^{ }_l - 
  Q_l^\dagger \, G^{lk} \, P + 
  \epsilon^{ }_{mln} Q^\dagger_m \, G^{lk} \, Q^{ }_n
 \Bigr]
 \nn & + &  
 i g_\pi 
 \Bigl[ 
  P^\dagger (G^{00} - i \Delta M \mathcal{A}^{0\,T} ) Q^{ }_k - 
  Q_k^\dagger (G^{00} + i \Delta M \mathcal{A}^{0\,T} )  P + 
  \epsilon^{ }_{mkn} Q^\dagger_m \, G^{00} \, Q^{ }_n
 \Bigr]
 \nn & + &  
 g_\pi^2 
 \Bigl[ 
  Q_k^\dagger \, \tilde{H}^{l0} \, Q^{ }_l - 
  Q_l^\dagger \, \tilde{H}^{l0} \, Q^{ }_k 
 + \epsilon^{ }_{klm}
 \Bigl(
  Q_m^\dagger \, \tilde{H}^{l0} \, P^{ } - 
  P^\dagger \, \tilde{H}^{l0} \, Q^{ }_m 
 \Bigr)
 \Bigr]
 \biggr\} + \rmO\Bigl( \frac{1}{M^2} \Bigr)
 \;, \nn \la{force}
\ea
where 
$
 \Delta M \equiv M^{ }_Q - M^{ }_P
$
and flavour indices have been suppressed. 
For $\Delta M = 0$
each row is separately invariant under  the heavy quark spin 
symmetry, \eq\nr{symm}. We focus on two  
of the operators in the following, namely 
$ 
 \sim P^\dagger F P
$ 
(cf.\ \eq\nr{nlo_force})
and 
$
 \sim i g_\pi P^\dagger G Q
$
(cf.\ \eq\nr{lo_sketch}). 

%
\subsection{Euclidean formulation}

Before proceeding with the main line of the computation, 
we need to confront the specific 
time ordering appearing in \eq\nr{kappa_def}. We do this by going 
through the imaginary-time formalism, because this 
produces an intermediate result which could in principle
be compared with lattice simulations. Denoting by $\tau = it$
the Euclidean time coordinate, the propagator of $P^{ }_a$ 
is then determined by the Euclidean action 
(corresponding to a weight $\exp(-S_E)$)
\be
 S_E^{(0)} = \int_0^\beta \! {\rm d}\tau \int_{\vec{x}}
 2\, P^\dagger_a ( \partial_\tau + M^{ }_P  ) P^{ }_a 
 + (P^{ }_a \leftrightarrow Q^{ }_{ak} )
 \;. \la{S_E}
\ee 
Due to charge conservation 
the susceptibility of \eq\nr{susc} can be re-expressed as
\be
 \chi^{00} = \int_0^\beta \! {\rm d}\tau \int_{\vec{x}}
 \langle
   \, [\, 2 P^\dagger_a P^{ }_a \,] (\tau,\vec{x})
   \, [\, 2 P^\dagger_b P^{ }_b \,] (0,\vec{0}) \,
 \rangle^{ }_T 
 + (P^{ }_a \leftrightarrow Q^{ }_{ak} )
 + \rmO\Bigl(\frac{1}{M} \Bigr)
 \;, \la{susc_E}
\ee
and, taking into account 
the Wick rotation of the time coordinate, the Euclidean correlator 
related to \eq\nr{kappa_def} can be defined as 
\ba
 G_\rmii{E}(\tau) & \equiv & 
 \frac{\beta}{3} 
  \lim_{M\to\infty} \frac{M^2}{\chi^{00}}
 \int_{\vec{x}} 
 \left. \langle
  \partial^0 \hat {\cal{J}}^k(\tau,\vec{x}) 
    \;
  \partial^0 \hat {\cal{J}}^k(0,\vec{0})    
 \rangle^{ }_T \right|_{it\to \tau}
 \;. 
 \la{Gtau_def} 
\ea
The subscript in $G_\rmii{E}$ could refer to ``electric''.
In the subsequent steps we again  
omit hats, because \eq\nr{Gtau_def} can be evaluated 
through normal Euclidean path integrals. 

The correlator of \eq\nr{Gtau_def} is periodic across the Euclidean
time interval, 
$
 G_\rmii{E}(\tau + k \beta) = G_\rmii{E}(\tau)
$, $k \in \mathbbm{Z}$, 
and can be Fourier analyzed in the Matsubara
formalism. In particular, after a Fourier transform,  
\be
  \tilde G_\rmii{E}(\omega_n) = 
  \int_0^\beta \! {\rm d}\tau \, e^{i \omega_n\tau } G_\rmii{E}(\tau)
  \;, \quad
  \la{Fourier}
\ee
where $\omega_n = 2\pi n T$, $n \in \mathbbm{Z}$, 
the spectral function is obtained from an imaginary part~\cite{ftft}:
\be
  \rho_\rmii{E}(\omega) = 
  \im \tilde G_\rmii{E}(\omega_n \to -i [\omega + i 0^+])
 \;. \la{spectral}
\ee
The momentum diffusion coefficient then follows from \eq\nr{relation}. 

The Euclidean action in \eq\nr{S_E} implies that, 
for $0 < | \tau - \sigma | < \beta$, 
the free propagator is 
\ba
  && \hspace*{-1cm} 
 \langle
   P^{ }_a(\tau,\vec{x}) \, P_b^\dagger(\sigma,\vec{y}) 
 \rangle^{(0)}_T
 \nn 
  & = &
   \frac{1}{2} \delta_{ab}\, \delta^{(3)}(\vec{x}-\vec{y})
     \, T \sum_{\omega_n} 
     \frac{e^{i\omega_n (\tau-\sigma)}}{i\omega_n + M^{ }_P}
  \nn 
  & = & 
  \frac{1}{2}\delta_{ab}\, \delta^{(3)}(\vec{x}-\vec{y}) 
  \, \nB{}(M^{ }_P) 
  \Bigl[
    \theta(\sigma-\tau) e^{(\sigma-\tau) M^{ }_P} +  
    \theta(\tau-\sigma) e^{(\beta - \tau + \sigma) M^{ }_P }
  \Bigr]
  \;,  \la{freeP}
\ea
where 
$\nB{}(M^{ }_P) \equiv 1/(e^{\beta M^{ }_P} -1)$ is the Bose distribution. 
The propagator 
$ 
 \langle Q^{ }_{ak} Q^\dagger_{bl} \rangle^{(0)}_T
$
has the same structure, with an additional $\delta_{kl}$.
The susceptibility of \eq\nr{susc_E} then evaluates to
\be
 \chi^{00} = 
 \beta \Nf \, \delta^{(3)}(\vec{0}) \, 
 \Bigl( e^{-\beta M^{ }_P} + 3\, e^{-\beta M^{ }_Q} \Bigr)
 + \rmO\Bigl(\frac{1}{F^2},\frac{1}{M}\Bigr)
 \;, \la{susc_res}
\ee
where we also took the limit $M^{ }_P, M^{ }_Q \gg T$, replacing 
$\nB{}(M^{ })$ through $\exp(-\beta M^{ })$.

%
\subsection{Leading order}

%
\begin{figure}[t]

\hspace*{0.5cm}%
\begin{minipage}[c]{4.5cm}
\begin{picture}(120,80)(-60,-40)
\SetScale{0.9}  
\SetWidth{1.5}
\CArc(0,0)(30,90,270)%
\CArc(0,0)(28.5,-90,90)%
\CArc(0,0)(31.5,-90,90)%
\DashLine(0,30)(0,-30){5}%
\GBoxc(0,30)(7,7){1}%
\GBoxc(0,-30)(7,7){1}%
\Text(-36,0)[c]{$P$}%
\Text(7,0)[c]{$\pi$}%
\Text(36,0)[c]{$Q$}%
\Text(0,-45)[c]{(a)}
\end{picture}
\end{minipage}%
\hspace*{0.5cm}%
\begin{minipage}[c]{4.5cm}
\begin{picture}(120,80)(-60,-40)
\SetScale{0.9}  
\SetWidth{1.5}
\CArc(0,0)(30,90,270)%
\CArc(0,0)(30,-90,90)%
\DashCArc(33,0)(44,137,223){5}%
\DashCArc(-33,0)(44,-43,43){5}%
\GBoxc(0,30)(7,7){1}%
\GBoxc(0,-30)(7,7){1}%
\Text(-36,0)[c]{$P$}%
\Text(-5,0)[c]{$\pi$}%
\Text(15,0)[c]{$\pi$}%
\Text(36,0)[c]{$P$}%
\Text(0,-45)[c]{(b)}
\end{picture}
\end{minipage}%
\hspace*{0.5cm}%
\begin{minipage}[c]{4.5cm}
\begin{picture}(120,80)(-60,-40)
\SetScale{0.9}  
\SetWidth{1.5}
\CArc(0,0)(30,90,180)%
\CArc(0,0)(30,-90,0)%
\CArc(0,0)(28.5,0,90)%
\CArc(0,0)(31.5,0,90)%
\CArc(0,0)(28.5,180,270)%
\CArc(0,0)(31.5,180,270)%
\DashLine(0,30)(0,-30){5}%
\DashLine(-31.5,0)(-5,0){5}%
\DashLine(5,0)(31.5,0){5}%
\GBoxc(0,30)(7,7){1}%
\GBoxc(0,-30)(7,7){1}%
\Text(-25,25)[c]{$P$}%
\Text(25,-25)[c]{$P$}%
\Text(6,15)[c]{$\pi$}%
\Text(-15,-6)[c]{$\pi$}%
\Text(25,25)[c]{$Q$}%
\Text(-25,-25)[c]{$Q$}%
\Text(0,-45)[c]{(c)}
\end{picture}
\end{minipage}%

\vspace*{0.7cm}

\caption[a]{\small 
Euclidean correlators corresponding to the scattering 
processes shown in \fig\ref{fig:Egraphs}. The solid 
(single or double-lined) circle represents the Euclidean 
time interval; open squares correspond to force operators; 
and $P$ and $Q$ stand for the pseudoscalar and vector fields, 
respectively (in particle language the $Q$-field is denoted by $P^*$).
The pion lines do {\em not} meet in process (c).
} 
\la{fig:Egraphs}
\end{figure}
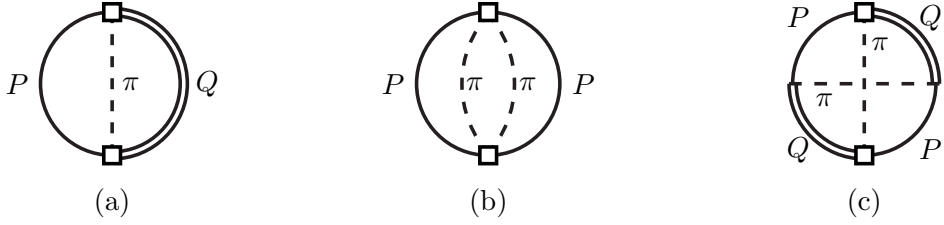
%

We now move on to consider the correlator
$G_\rmii{E}$, defined by \eq\nr{Gtau_def}, with the force
inserted from \eq\nr{force} and the susceptibility from 
\eq\nr{susc_res}. Some of the relevant Feynman graphs are 
shown in \fig\ref{fig:Egraphs}; we start by considering
the process (a). It was already argued in connection with
\fig\ref{fig:graphs} that this should give no 
contribution to $\kappa$ but, as a preparation for 
the next-to-leading order computation, we 
recall briefly how the vanishing can be seen. 

Making use of \eqs\nr{lo_VA}, \nr{G_def} and the 2nd and 3rd 
rows of \eq\nr{force}, 
the process in \fig\ref{fig:Egraphs}(a) corresponds to 
\be
 \langle 
  \, \partial^0 \mathcal{J}_k (\tau,\vec{x})
  \, \partial^0 \mathcal{J}_k (0,\vec{0}) \,
 \rangle^{ }_T \sim  
 \frac{g_\pi^2}{M^2 F^2}
 \Bigl\langle 
   [\, Q^\dagger_a \; \partial \partial \xi_{ba} \, P_b \,]
    (\tau,\vec{x}) \;
   [\, P^\dagger_c \; \partial \partial \xi_{dc} \, Q_d \,]
    (0,\vec{0})
 \Bigr\rangle^{ }_T
 \;, \la{lo_sketch}
\ee
where space-time indices have been omitted. 
Contracting the heavy-meson fields as in \eq\nr{freeP}, 
we note that there is one appearance of ``forward'' 
and one of ``backward'' propagation; in the imaginary-time 
formalism, these amount to circling the Euclidean time interval. 
If we also insert the Goldstone boson propagator from \eq\nr{G_prop}, 
the result becomes  
\ba
 && \hspace*{-2cm}
 \int_{\vec{x}}
  \langle 
  \, \partial^0 \mathcal{J}_k (\tau,\vec{x})
  \, \partial^0 \mathcal{J}_k (0,\vec{0}) \,
 \rangle^{ }_T 
 \nn 
 & \sim &
 \frac{g_\pi^2 (\Nf^2 - 1)}{M^2 F^2} \, \delta^{(3)}(\vec{0})
 \,
 \nB{}(M^{ }_P)
 \,
 \nB{}(M^{ }_Q) 
 \,
 e^{(\beta-\tau) M^{ }_P}
 \,
 e^{\tau M^{ }_Q} 
 \,
 \partial\partial\partial\partial \Delta(\tau,\vec{0})    
 \;, \la{sketch}
\ea
where 
\be
 \Delta(\tau,\vec{0}) 
 = T \sum_{\omega_n} \int_{\vec{p}} 
  \frac{e^{i \omega_n\tau + i \vec{p}\cdot\vec{0}}}
  {\omega_n^2 + E_p^2}
 = \int_{\vec{p}} 
 \frac{\nB{}(E_p)}{2 E_p}
 \Bigl[
    e^{\tau E_p} + e^{(\beta-\tau)E_p} 
 \Bigr] e^{i \vec{p}\cdot\vec{0}}
 \;, \la{Delta_def}
\ee
with $E_p \equiv \sqrt{m_\pi^2 + p^2}$ and
$\int_{\vec{p}} \equiv \int  {\rm d}^3\vec{p} / (2\pi)^3$.
A Fourier transform (cf.\ \eq\nr{Fourier}) can now be taken: 
\ba
 && \hspace*{-2cm}
 \int_0^\beta \! {\rm d}\tau \, e^{i \omega_n\tau } 
 e^{(\beta-\tau) M^{ }_P}
 e^{\tau M^{ }_Q} 
 \Bigl[
    e^{\tau E_p} \pm e^{(\beta-\tau)E_p} 
 \Bigr]
 \nn 
 & = & \frac{e^{\beta(M^{ }_Q + E_p)} - e^{\beta M^{ }_P}}
   {i\omega_n + M^{ }_Q - M^{ }_P + E_p} \pm 
       \frac{e^{\beta M^{ }_Q} - e^{\beta ( M^{ }_P + E_p )}}
   {i\omega_n + M^{ }_Q - M^{ }_P - E_p}
 \;. \la{lo_iwn}
\ea
The subsequent discontinuity, \eq\nr{spectral}, turns these 
into energy conservation constraints: 
\ba
 \rho_\rmii{E}(\omega)
 \!\! & \propto & \!\! 
 \int_{\vec{p}} \im \biggl[ 
   \frac{e^{\beta(M^{ }_Q + E_p)} - e^{\beta M^{ }_P}}
   {\omega + M^{ }_Q - M^{ }_P + E_p + i 0^+} \pm 
       \frac{e^{\beta M^{ }_Q} - e^{\beta ( M^{ }_P + E_p )}}
   {\omega + M^{ }_Q - M^{ }_P - E_p + i 0^+}
 \biggr]
 \nn 
 & = & -\pi \int_{\vec{p}} \Bigl[ 
    \delta(\omega + M^{ }_Q - M^{ }_P + E_p)  e^{\beta E_p}
 \pm  \delta(\omega + M^{ }_Q - M^{ }_P - E_p)
 \Bigr] e^{\beta M^{ }_Q} \Bigl(  1 - e^{\beta\omega} \Bigr)
 \;, \nn \la{lo_rho}
\ea
where the $\delta$-functions were made use of for 
rewriting the exponentials. (Using Hermitean conjugate
operators in \eq\nr{lo_sketch} yields the same but 
with $M^{ }_Q \leftrightarrow M^{ }_P$.) 

It can now be seen that the limit of \eq\nr{relation} 
can be non-trivial only if there are pion
momenta with $E_p = \pm (M^{ }_Q - M^{ }_P)$
(in fact this is a necessary but not a sufficient condition; in principle 
it could happen that the numerator of \eq\nr{sketch}, which has 
been left implicit here, vanishes at the same point, however 
this does not appear to be the case). 
This corresponds indeed to \fig\ref{fig:graphs}(a), and cannot 
be realized if $|M^{ }_Q - M^{ }_P| < m_\pi$,
as is the case with $B$-mesons. 
Let us end by remarking that the same is expected to be 
the case also for many corrections of $\rmO(g_\pi^2/F^4)$, 
obtained from the topology of \fig\ref{fig:Egraphs}(a) by 
dressing either the pion propagator or the force operator 
by a closed ``bubble'', which does not affect momentum flow.

%
\subsection{Next-to-leading order}

Moving on to next-to-leading order, we need to consider graphs 
like (b) and (c) in \fig\ref{fig:Egraphs}. Due to the large number
of terms proportional to $g_\pi$ in \eq\nr{force}, as well as the possibly
associated uncertainty as discussed below \eq\nr{symm}, we simplify 
the task here by setting $g_\pi\to 0$, omitting thereby contributions
of $\rmO(g_\pi^2)$ and $\rmO(g_\pi^4)$ from $\kappa$, which would otherwise
be of the same order in the chiral expansion. Recalling that 
phenomenologically $g_\pi \simeq 0.5$, this cannot change
the overall magnitude of the result. In this limit only 
the process (b) 
(as well as the same topology with $P^{ }_a \to Q^{ }_{ak}$) 
is left over.
Apart from ``trivial'' mass effects, spin symmetry implies 
that the terms containing $P_a$ and $Q_{ak}$ 
have identical structures at $\rmO(g_\pi^0)$.  Therefore 
in the following only the part involving $P_a$ 
is displayed explicitly.  

As a first step we rewrite the relevant term of \eq\nr{force}: 
\be
 \partial^0 \int_{\vec{x}} \mathcal{J}^k 
 = \frac{2i}{M} 
 \int_{\vec{x}} P_a^\dagger ( \mathcal{D}^0_{ca} \mathcal{D}^k_{bc} 
                      - \mathcal{D}^k_{ca} \mathcal{D}^0_{bc} ) P^{ }_b
 + \rmO\Bigl(\frac{g_\pi}{M} \Bigr)
 \;. \la{nlo_force}
\ee
This can be contrasted with \eq(2.17) of ref.~\cite{eucl}: 
the colour-electric field strength of HQET has been replaced 
by a kind of ``chiral-electric'' field strength in HM\XPT{}. 

Going then over to the Euclidean correlator of \eq\nr{Gtau_def}, 
we obtain
\ba
 G_\rmii{E}(\tau)  & = & 
 \frac{\beta}{3} 
 \lim_{M\to\infty} \frac{4}{\chi^{00}}
 \int_{\vec{x}}
  \Bigl\langle
  \Bigl\{
          P_a^\dagger [ \mathcal{D}^T_\tau , \mathcal{D}^T_{k} ]^{ }_{ab} 
                       P^{ }_b 
  \Bigr\} (\tau,\vec{x}) \,
  \Bigl\{
          P_c^\dagger [ \mathcal{D}^T_\tau , \mathcal{D}^T_{k} ]^{ }_{cd} 
                       P^{ }_d 
  \Bigr\} (0,\vec{0})
 \Bigr\rangle_T
  + \rmO\bigl( g_\pi^2 \bigr)
 \;. \nn \la{Gtau}
\ea
Inserting the heavy meson propagator from \eq\nr{freeP}
and the susceptibility from \eq\nr{susc_res}, this can 
be re-expressed as
\be
  G_\rmii{E}(\tau) = 
 \frac{1}{3 \Nf} 
  \Bigl\langle
                [ \mathcal{D}^T_\tau , \mathcal{D}^T_{k} ]^{ }_{ab} 
  (\tau,\vec{0}) \,
                [ \mathcal{D}^T_\tau , \mathcal{D}^T_{k} ]^{ }_{ba} 
  (0,\vec{0})
 \Bigr\rangle_T
  + \rmO\Bigl( \frac{g_\pi^2}{F^4}, \frac{1}{F^6} \Bigr)
 \;. \la{Gtau_2}
\ee
With the contribution of 
$Q^{ }_{ak}$ added as has already been done here, the Boltzmann weights
(cf.\ \eq\nr{susc_res}) have 
duly cancelled between the numerator and the denominator. 

Having obtained \eq\nr{Gtau_2}, 
the problem has reduced to one within normal 
Chiral Perturbation Theory, \eq\nr{L_XPT}. Noting from 
\eqs\nr{lo_VA}, \nr{F_def} that 
\ba
  [ \mathcal{D}^T_\tau , \mathcal{D}^T_{k} ]^{ }_{ab}
 & = & 
  i ( \partial_\tau \mathcal{V}_k - \partial_k \mathcal{V}_\tau )^{ }_{ba} 
  + \rmO\Bigl(\frac{1}{F^4} \Bigr)
 \\  
 & = &  
   \frac{1}{F^2} (\partial_k \xi \, \partial_\tau \xi - 
                  \partial_\tau \xi \, \partial_k \xi)^{ }_{ba}
   + \rmO\Bigl(\frac{1}{F^4} \Bigr)
   \;, 
\ea
and inserting the propagator from \eq\nr{G_prop}, 
we obtain after some algebra that 
\be
 G_\rmii{E}(\tau) = \frac{\Nf^2-1}{6 F^4}
 \Bigl[ 
   \partial_\tau \partial_k \Delta(\tau,\vec{0})
   \partial_\tau \partial_k \Delta(\tau,\vec{0}) - 
   \partial_\tau^2 \Delta(\tau,\vec{0}) 
   \nabla^2 \Delta(\tau,\vec{0})
 \Bigr]
  + \rmO\Bigl( \frac{g_\pi^2}{F^4}, \frac{1}{F^6} \Bigr)
 \;. \la{Gtau_3}
\ee
Employing $\Delta(\tau,\vec{0})$ from \eq\nr{Delta_def},
the first term of \eq\nr{Gtau_3}
is seen not to contribute, because the $\vec{p}$-integrand
is odd in $\vec{p}\to -\vec{p}$. 
The second term of \eq\nr{Gtau_3} does contribute; 
the Fourier transform (cf.\ \eq\nr{Fourier}) can be carried out 
and, apart from a contact term $\propto \; \delta(\tau)$, 
yields a sum of four terms, with $\omega_n$-dependence 
in structures of the type $\sim 1/(i\omega_n \pm E_p \pm E_q)$, 
in analogy with \eq\nr{lo_iwn}. 
The spectral function (cf.\ \eq\nr{spectral}) is obtained by replacing
these with $-\pi \delta(\omega  \pm E_p \pm E_q)$, 
in analogy with \eq\nr{lo_rho}. 
Thereby we end up with
\ba
 \rho_\rmii{E}(\omega) & = &  \frac{\pi(\Nf^2-1)}{6 F^4}
 \int_{\vec{p},\vec{q}} \frac{p^2 E_q}{4 E_p}
 \nn & & \times \, 
 \Bigl\{
   \Bigl[ 1 + \nB{}(E_p) + \nB{}(E_q) \Bigr]
   \Bigl[\delta(\omega-E_p-E_q) - \delta(\omega + E_p + E_q)\Bigr] 
 \nn & & \hspace*{5mm} + \, 
   \Bigl[ \nB{}(E_p) - \nB{}(E_q) \Bigr]
   \Bigl[\delta(\omega+E_p-E_q) - \delta(\omega - E_p + E_q) \Bigr] 
 \Bigr\}
 \nn & + & 
  \rmO\Bigl( \frac{g_\pi^2}{F^4}, \frac{1}{F^6} \Bigr)
 \;. \la{rho_res}
\ea

The final step is to take the limit defined in \eq\nr{relation}.
Due to vanishing phase space, only the second structure of 
\eq\nr{rho_res} gives a contribution linear in $\omega$ at $\omega\to 0$; 
this corresponds to the process in \fig\ref{fig:graphs}(b) 
in the limit of vanishing energy transfer. 
The terms linear in $\omega$ can be extracted 
by using the $\delta$-functions to rewrite the arguments of the Bose
distributions, and by then Taylor-expanding the latter. There are
two terms, amounting to a symmetrization $\vec{p}\leftrightarrow \vec{q}$; 
afterwards we can set $E_q \to E_p$. Given that 
$
 \nB{}'(E) = -\beta \nB{}(E)[1+\nB{}(E)]
$, 
this yields
\ba
 \kappa & = &
 \frac{\pi(\Nf^2-1)}{3 F^4}
 \int_{\vec{p},\vec{q}} \frac{p^2 + q^2}{4}
   \nB{}(E_p) 
   \Bigl[ 1 + \nB{}(E_p) \Bigr]
   \delta(E_p-E_q)
  + \rmO\Bigl(\frac{g_\pi^2}{F^4},\frac{1}{F^6} \Bigr)
 \nn & = & 
 \frac{(\Nf^2-1) T}{24 \pi^3 F^4}
 \int_0^\infty \! {\rm d}p^2 \, p^2 \, (3 p^2 + 2 m_\pi^2)  \, \nB{}(E_p) 
  + \rmO\Bigl(\frac{g_\pi^2}{F^4},\frac{1}{F^6} \Bigr)
 \;, \la{kappa_final}
\ea
where a partial integration was carried out. 
In the limit $\pi T \gg m_\pi$ the integral can be performed explicitly
and the result reads
\be
 \kappa \; \stackrel{\pi T \gg m_\pi}{\approx} \;
 \frac{2(\Nf^2-1)\pi^3 T^7}{63 F^4}
 \biggl[
  1 - \frac{7 m_\pi^2}{10 \pi^2 T^2}
    + \rmO \Bigl( \frac{m_\pi^4}{\pi^4 T^4} \Bigr) 
 \biggr] +
 \rmO\biggl(\frac{g_\pi^2 T^7}{F^4},\frac{T^9}{F^6} \biggr)
 \;. \la{kappa_asympt}
\ee
A numerical evaluation is shown in \fig\ref{fig:2}.
This constitutes our final result. 

\begin{figure}[t]

\centerline{%
 \epsfysize=7.5cm\epsfbox{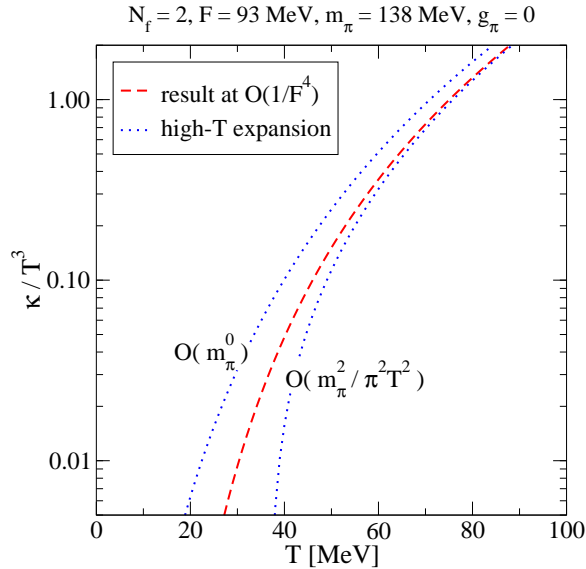}%
}
 
\caption[a]{\small 
 An illustration of $\kappa/T^3$ as a function
 of the temperature. 
 The dashed line shows the result from 
 \eq\nr{kappa_final}; the dotted lines show two orders of the 
 high-temperature expansion from  
 \eq\nr{kappa_asympt}. At high temperatures, $T \gsim 80$~MeV, 
 the chiral expansion breaks down because higher order 
 corrections of relative magnitude $\rmO(T^2/F^2)$ become 
 large; at low temperatures, $T \lsim 30$~MeV, the high-temperature
 expansion breaks down because the expansion parameter 
 $\rmO(m_\pi^2/\pi^2 T^2)$ exceeds unity (the unexpanded result 
 of \eq\nr{kappa_final} remains valid but is exponentially small). 
 In the deconfined phase
 a value $\kappa/T^3 \gsim 2$ might be  
 phenomenologically acceptable (see e.g.\ ref.~\cite{ct}; 
 an oft-cited ``diffusion coefficient'' 
 is $D \simeq 2 T^2/\kappa \lsim 1/T$).
 The weak-coupling expansion suggests in general 
 values $\kappa/T^3 \lsim 2$~\cite{chm}, and at very high 
 temperatures $\kappa/T^3$ decreases like $\sim 1/\ln^2(T/T_0)$.  
 }
\la{fig:2}
\end{figure}

%
\section{Conclusions and outlook}
\la{se:concl}

The purpose of this paper has been to estimate the overall
magnitude of the momentum 
diffusion coefficient, $\kappa$, of a heavy-light pseudoscalar 
meson almost at rest with respect to a heat bath, 
which has a temperature in the range of some tens of MeV. 
The result, \eq\nr{kappa_final} and \fig\ref{fig:2}, is given 
in terms of low-energy constants of two-flavour 
Chiral Perturbation Theory. 

Although the result obtained is
not immediately applicable to heavy ion collision experiments, 
in which the initial temperature may be in the range of hundreds 
of MeV and much of heavy quark energy loss
could take place in the deconfined phase, the hope is that the analysis
is nevertheless of theoretical interest. After all, 
physical QCD is believed to possess a crossover rather than a genuine
phase transition between low and high temperatures, so that analyses
in the former regime may yield  qualitative information
also for the latter. More concretely, our analysis 
combined with existing weak-coupling computations~\cite{chm} suggest 
a picture in which the momentum diffusion coefficient exhibits a broad peak 
around the QCD crossover. So, heavy quark jets produced in 
an initial hard process may continue to approach kinetic equilibrium 
all the way until their final decoupling within the hadronic phase.  

Despite the qualitative nature of our study, it seems that 
a number of potentially interesting extensions can be envisaged. 
Most obviously, the result for $\kappa$ of $B$-mesons could be 
completed with terms of $\rmO(g_\pi^2)$ and $\rmO(g_\pi^4)$. 
Although no qualitative changes are expected, 
the corrections may be numerically important, perhaps
in the range of $\sim 50$\%. In addition the analysis appears formally 
interesting: as discussed in the text, it should probably be carried 
out both within the non-relativistic theory as well as within 
a relativistic extension thereof, taking the non-relativistic limit 
only afterwards in the latter case, in order to have a crosscheck. 

Another interesting topic is the ``extension'' of the study to $D$-mesons. 
This immediately leads to the dramatic effect that resonant contributions, 
such as the process shown in \fig\ref{fig:graphs}(a), are allowed, 
whereby the $D$-mesons can change their identity
(a further complication is that for $D^*$ decays electromagnetic 
processes are important). The handling of the related 
rich physics may suggest an excursion away from
systematic computations, into 
the realm of hadronic models; an overview of recent works
in this direction can be found e.g.\ in ref.~\cite{D1}. 
Nevertheless, it can be noted that resonant phenomena 
could also play a role in theoretical considerations 
of the $B$-sector, if we took the chiral limit $m_\pi\to 0$ while keeping 
the mass difference $M^{ }_{B^*} - M^{ }_B$ fixed and non-zero. 

Yet another line 
is that whereas only the ``intercept'', 
$
 \kappa = \lim_{\omega\to 0} 2 T \rho_\rmii{E}(\omega)/\omega
$, 
was addressed here, 
the full spectral function $\rho_\rmii{E}(\omega)$ (\eq\nr{spectral}), 
and even the Euclidean correlator $G_\rmii{E}(\tau)$ (\eq\nr{Gtau_def}), 
are also interesting objects. For instance,  
they may help in the interpretation
of the corresponding lattice measurements {\it \`a la} ref.~\cite{hbm2}
(if these were unquenched);
it is in this spirit that both functions have been computed also
in the deconfined phase~\cite{rhoE}, and in fact $G_\rmii{E}(\tau)$
even in the confined phase in the presence of a large lattice 
spacing, through the use of the lattice 
strong-coupling expansion~\cite{rhoE}. 
The terms of $\rmO(g_\pi^2/F^2)$
(cf.\ \fig\ref{fig:Egraphs}(a)) {\em do} contribute 
significantly to $G_\rmii{E}(\tau)$, even if they do not contribute 
to $\kappa$ as has been discussed in the text; 
it might turn out to be useful to gain understanding on how hidden
the information about $\kappa$ is in the directly 
measurable $G_\rmii{E}(\tau)$.

A further topic is not to consider 
the spectral function $\rho_\rmii{E}(\omega)$ yielding $\kappa$, but rather
the {\em full} spectral function 
$
 \rho(\omega,\vec{k})
 \equiv 
 \int_{t,\vec{x}} e^{i\omega t - i \vec{k}\cdot\vec{x}}
 \langle \fr12 [ 
  \hat{\mathcal{J}}^\mu (t,\vec{x}) , 
  \hat{\mathcal{J}}_\mu (0,\vec{0})
  ] 
 \rangle^{ }_T
$, 
of which $\rho_\rmii{E}(\omega)$ 
is a specific limit~\cite{eucl}. 
Restricting first to vanishing momentum, $\vec{k} = \vec{0}$, and 
to small frequencies, the analysis can still be 
carried out within the non-relativistic framework. This spectral
function is expected to show a narrow and prominent ``transport peak''
at frequencies $|\omega| \lsim \eta_\rmii{D} \sim \kappa / (2\, T M)$. 
A direct analysis of such infrared features tends to be difficult, 
necessitating complicated resummations (see e.g.\  ref.~\cite{lgdm}), 
but at least the theory in question is not a gauge theory, whereby these
may be more tractable than in the deconfined phase. 

The most ambitious goal would be to consider $\rho(\omega,\vec{k})$
also for $\vec{k}\neq \vec{0}$, 
perhaps even for $|\vec{k}| \gsim M$, corresponding
to heavy mesons moving at a relativistic speed with respect to 
a pionic plasma. Although this could take us away from the range
of validity of systematic HM\XPT,\ simple relativistic extensions
can be written down~(see e.g.\ refs.~\cite{invM,bj}). Within 
such a framework one might also try to understand 
phenomena such as radiative energy loss, 
perhaps making contact with classic pion gas computations that 
have recently been revived through the AdS/CFT setup~\cite{cfm}.

%
\section*{Acknowledgements}

This work was partly supported by the BMBF under project
{\em Heavy Quarks as a Bridge between
     Heavy Ion Collisions and QCD}. 

%
\section*{Note added}

After the submission of this paper three works 
appeared~\cite{He:2011yi}--\cite{Abreu:2011ic} 
in which the kinetic
equilibration of $D$-mesons is studied in a spirit similar
to that suggested in \se\ref{se:concl}.


\appendix
\renewcommand{\thesection}{Appendix~\Alph{section}}
\renewcommand{\thesubsection}{\Alph{section}.\arabic{subsection}}
\renewcommand{\theequation}{\Alph{section}.\arabic{equation}}


\end{document}